\documentclass[12pt]{iopart}

\usepackage{epsfig}
\usepackage{color}
\newcommand{\ful}{\mbox{C$_{\mbox{\scriptsize{60}}}$}}
\newcommand{\vscf}{\mbox{$\delta V$}}
\newcommand{\eq}[1]{Eq.~\ref{#1}}

\begin{document}
\title[DFT study of photoionization ICD resonances in Cl@$\ful$]
{Density functional study of the variants of inter-Coulombic decay resonances in the photoionization of Cl@$\ful$}

\author{Ruma De$^1$, Esam Ali$^1$, Steven T Manson$^2$, and Himadri S Chakraborty$^{1}$}

\address{$^1$Department of Natural Sciences, D. L. Hubbard Center for Innovation and Entrepreneurship, Northwest Missouri 
State University, Maryville, Missouri 64468, USA}
\address{$^2$Department of Physics and Astronomy, Georgia State University, Atlanta, Georgia, USA}

\ead{himadri@nwmissouri.edu}

\begin{abstract}
Inter-Coulombic decay (ICD) resonances in the photoionization of Cl@$\ful$ endofullerene molecule are calculated  using a perturbative density functional theory (DFT) method. This is the first ICD study of an open shell atom in a fullerene cage. Three classes of resonances are probed: (i) Cl inner vacancies decaying through $\ful$ outer continua, (ii) $\ful$ inner vacancies decaying through Cl outer continua, and (iii) inner vacancies of either system decaying through the continua of Cl-$\ful$ hybrid levels, the hybrid Auger-ICD resonances. Comparisons with Ar@$\ful$ results reveal that the properties of hybrid Auger-ICD resonances are affected by the extent of level hybridization.

\end{abstract}


\newpage

\section{Introduction}

In loosely bound composite matters, such as polymers, liquids, and biological systems, the relaxation of an innershell vacancy resulting in the emission of an outershell electron, both belonging to the same site of the system, is the regular Auger process. But, this vacancy can also decay by transferring excess energy to a neighboring site. This migrated energy can subsequently drive the emission of an electron from that site. Such processes, the inter-Coulombic decay (ICD), are abundant in nature when energetically allowed, unless quenched by a competing process, and piggyback on the long range electron-electron Coulomb interactions. Broadly speaking, the excess energy-transfer to a neighboring site can be triggered {\em via} three distinct mechanisms. (i) An outer electron of the vacancy site can itself fill in the vacancy - the regular ICD~\cite{ouchi2011}. (ii) A weakly bound electron from the ionizing neighboring site can transfer and fill the vacancy - the electron transfer mediated decay (ETMD)~\cite{foerstel2011}. (iii) A slow passerby electron can be captured into the vacancy - the inter-Coulombic electron capture (ICEC)~\cite{sisourat2018}. Experimentally, the precursor excitation process to create the vacancy can be induced in varieties of ways: The early work of the observation of ICD in Ne dimers used synchrotron radiation for this purpose~\cite{jahnke2004rareDimer}. To achieve a higher pulse rate, for instance to carry out time-resolved experiment, free electron laser sources are more appropriate~\cite{schnorr2013}. Furthermore, charged particle impact, such as pulsed electron guns~\cite{yan2018} or alpha-particle impact~\cite{kim2011} have also been used. ICD signatures are probed by traditional methods of electron~\cite{marburger2003firstExp} and ion~\cite{wiegandt2019} spectroscopy, including various coincidence techniques~\cite{laforge2019}. Access to time-resolved ICD dynamics has also been possible by the contemporary pump-probe approaches~\cite{schnorr2013,takanashi2017}, specifically, by light field streaking techniques~\cite{trinter2013}. A recent comprehensive review of the experimental and theoretical research of the ICD topic, including the range of materials studied and potential applications, can be found in Ref.\ \cite{jahnke2020}.  
 
Probing ICD processes in relatively simpler vapor-phase materials is of considerable spectroscopic interest~\cite{barth2005ricdExp1,aoto2006ricdExp2,kim2013DimExp,trinter13HeNe,najjari10TwoCenter}. One class of such systems of current theoretical and experimental study is endofullerene complexes, in which an atomic or a small molecular host is placed in a fullerene cage. These are unique heterogeneous, nested dimers of weak host-fullerene bonding. From the experimental side, the synthesis techniques for these materials are fast-developing~\cite{popov2013endoSynth} with an advantage of their room-temperature stability. Furthermore, these materials are relevant in a number of applied contexts~\cite{popov2017}. And, note that measurements of a strong ICD signal in a molecular endofullerene, Ho$_3$N@C$_{80}$, has recently been reported~\cite{obaid2020}. 

If the electron that creates the vacancy subsequently fills the hole to release energy, the process is conventionally called the {\em participant} ICD. The first {\em ab initio} calculations of participant ICD induced resonances in the photoionization of $\ful$ levels induced by Ar inner $3s$~\cite{javani2014-rhaicd} and Kr inner $4s$~\cite{magrakvelidze2016} vacancy decays, the atom-to-fullerene ICD, were performed by our group. Later we also studied ICD resonances in the reverse process of fullerene-to-atom decay~\cite{de2016}. In addition, a remarkable coherence between the Auger and ICD amplitudes to produce a novel class of resonances in the photoionization of atom-fullerene hybridized states was also predicted~\cite{javani2014-rhaicd,magrakvelidze2016}. However, these studies cover only close-shell confined atoms. On the other hand, consideration of open-shell atomic endofullerenes to access their ICD properties arouses particular interest given their recent photoresponse studies~\cite{shields2020-jpb,shields2020-epjd}. In general, due to the existence of unpaired electrons, there are attractive fundamental interests in such systems. These include long spin relaxation times in N@$\ful$~\cite{morton2007} while enhancement and diminution in hyperfine coupling, respectively, in P@$\ful$~\cite{knapp2011} and exotic muonium@$\ful$~\cite{donzelli1996}. In this article, therefore, a prototypical open-shell system of Cl@$\ful$ has been considered for the first time to capture its ICD processes along the photoionization route. A comparison with the results of Ar@$\ful$, the nearest close-shell system of Cl@$\ful$, exposes the role of atom-$\ful$ hybridization in the Auger-ICD coherence process. 

\section{A fleeting description of theory}

Kohn-Sham density functional theory (DFT) is used to describe the ground, photoexcited, and photoionized electronic properties of Cl@$\ful$~\cite{shields2020-jpb}. The $\ful$ molecule is modeled by smudging sixty C$^{4+}$ ions over a classical spherical jellium shell, fixed in space, with an experimentally known $\ful$ mean radius of 3.5 \AA\ and thickness $\Delta$. The nucleus of a Cl atom is placed at the center of the sphere. The Kohn-Sham equations for the system of a total of $240+N$ electrons ($N=17$ for Cl and 240 delocalized electrons from $\ful$) are then solved to obtain the ground state properties in DFT. The gradient-corrected Leeuwen and Baerends exchange-correlation (XC) functional [LB94]~\cite{van1994exchange} is used for the accurate asymptotic behavior of the ground state radial potential
\begin{equation}\label{lda-pot}
V_{\scriptsize \mbox{DFT}}(\mathbf{r}) = V_{\mbox{\scriptsize jel}}({\bf r}) - \frac{z_{\mbox{\tiny atom}}}{r} + \int d\mathbf{r}'\frac{\rho(\mathbf{r}')}{|\mathbf{r}-\mathbf{r}'|} + V_{\scriptsize \mbox{\scriptsize XC}}[\rho(\mathbf{r})],
\end{equation} 
which is solved self-consistently in a mean-field framework. The requirement of charge neutrality produced $\Delta =$ 1.3 \AA, in agreement with the value inferred from experiment~\cite{ruedel2002oscExp,magrakvelidze2015}.

Linear-response time-dependent density functional theory (LR-TDDFT) is employed to simulate the dynamical response of $\ful$ to incident photons~\cite{madjet-jpb-08}. The single-electron dipole operator, $z$, corresponding to light that is linearly polarized in $z$-direction, induces a frequency-dependent complex change in the electron density arising from dynamical electron correlations. This can be written, using the independent particle (IP) susceptibility $\chi_0$, as
\begin{equation}\label{ind_den2}
\delta\rho({\bf r};\omega)={\int \chi_0({\bf r},{\bf r}';\omega) 
                           [z' + \vscf({\bf r}';\omega)] d{\bf r}'},
\end{equation}
in which
\begin{equation}\label{v_scf}
\vscf({\bf r};\omega) = \int\frac{\delta\rho({\bf r}';\omega)} {\left|{\bf r}-{\bf r}'\right|}d{\bf r}'
     +\left[\frac{\partial V_{\mbox{xc}}}{\partial \rho}\right]_{\rho=\rho_{0}} \!\!\!\!\delta\rho({\bf r};\omega),
\end{equation}
where the first and second terms on the right hand side are, respectively, the induced changes of the Coulomb and the exchange-correlation potentials. Obviously, $\vscf$ includes the dynamical field produced by important electron correlations within the linear response regime. In this method, the photoionization cross section corresponding to a bound-to-continuum dipole transition $n\ell\rightarrow k\ell^\prime$ is given by
\begin{equation}\label{cross-pi}
\sigma_{n\ell\rightarrow k\ell'} \sim |{\cal M}|^2 = |\langle k\ell'|z+\vscf|n\ell\rangle|^2,
\end{equation}
where, in the LR-TDDFT matrix element ${\cal M}$,  ${\cal D}$ and $\langle k\ell'|\vscf|n\ell\rangle$ are, respectively, the IP and correlation matrix elements. For the convenience of notation, we use the symbol $n\ell$@ to denote pure levels of the confined Cl atom and @$n\ell$ to represent pure levels of the doped $\ful$. 

In general, the full matrix element of photoionization of a level of Cl@$\ful$ can be written as:
\begin{eqnarray}\label{gen-mat-element}
{\cal M} (E) &=& {\cal D} (E) + {M}^{c-c} (E)+{M}^{d-c} (E),
\end{eqnarray}
where ${M}^{c-c}$ and ${M}^{d-c}$ are, respectively, contributions from continuum-continuum (c-c) and discrete-continuum (d-c) channel couplings. $\langle k\ell'|\vscf|n\ell\rangle$ in \eq{cross-pi} accounts for these coupling contributions. ${M}^{c-c}$ constitutes a rather smooth many-body contribution to nonresonant cross section, while the Auger or ICD resonances originate from ${M}^{d-c}$. 

\section{Results and discussions}

\subsection{Cl-to-$\ful$ ICD resonances}

Using the well-known approach by Fano~\cite{fano1961} to describe the dynamical correlation through the interchannel coupling, the amplitude of resonant ICD of Cl inner $3s@$ photo-vacancies {\em via} $\ful$ $@nl$ ionization can be expressed by $M^{\mbox{\scriptsize d-c}}$ that denotes the coupling of Cl $3s@\rightarrow \eta p@$ discrete excitation channels with the $@nl\rightarrow kl'$ continuum channel of $\ful$. Following~\cite{javani2014-rhaicd}, $M^{\mbox{\scriptsize d-c}}$ can thus be written as:
\begin{eqnarray}\label{dc-mat-element1}
 {M}^{\mbox{\scriptsize d-c}}_{@nl}(E) &=& \displaystyle\sum_{\eta}\sum_{l'} \frac{\langle\psi_{3s@ \rightarrow\eta p@}|\frac{1}{|{\bf r}_1-{\bf r}_2|}|\psi_{@nl\rightarrow kl'}(E)\rangle}{E-E_{3s@ \rightarrow\eta p@}} {\cal D}_{3s@ \rightarrow\eta p@},
\end{eqnarray}
where $E_{3s@ \rightarrow\eta p@}$ and ${\cal D}_{3s@ \rightarrow\eta p@}$ are, respectively, excitation energies and IP matrix elements of channels $3s@ \rightarrow\eta p@$, and $E$ is the photon energy corresponding to the $@nl$ transition to continuum. In \eq{dc-mat-element1} the $\psi$ are IP wavefunctions that represent the final states (channels) for transitions to excited $\eta p$ or continuum $kl'$ states. Obviously, the Coulomb coupling matrix element in the numerator of \eq{dc-mat-element1} acts as the passageway for energy transfer from the Cl de-excitation across to the $\ful$ ionization process, producing ICD resonances in the $\ful$ @$nl$ cross sections.
\begin{figure}
	\centerline{\psfig{figure=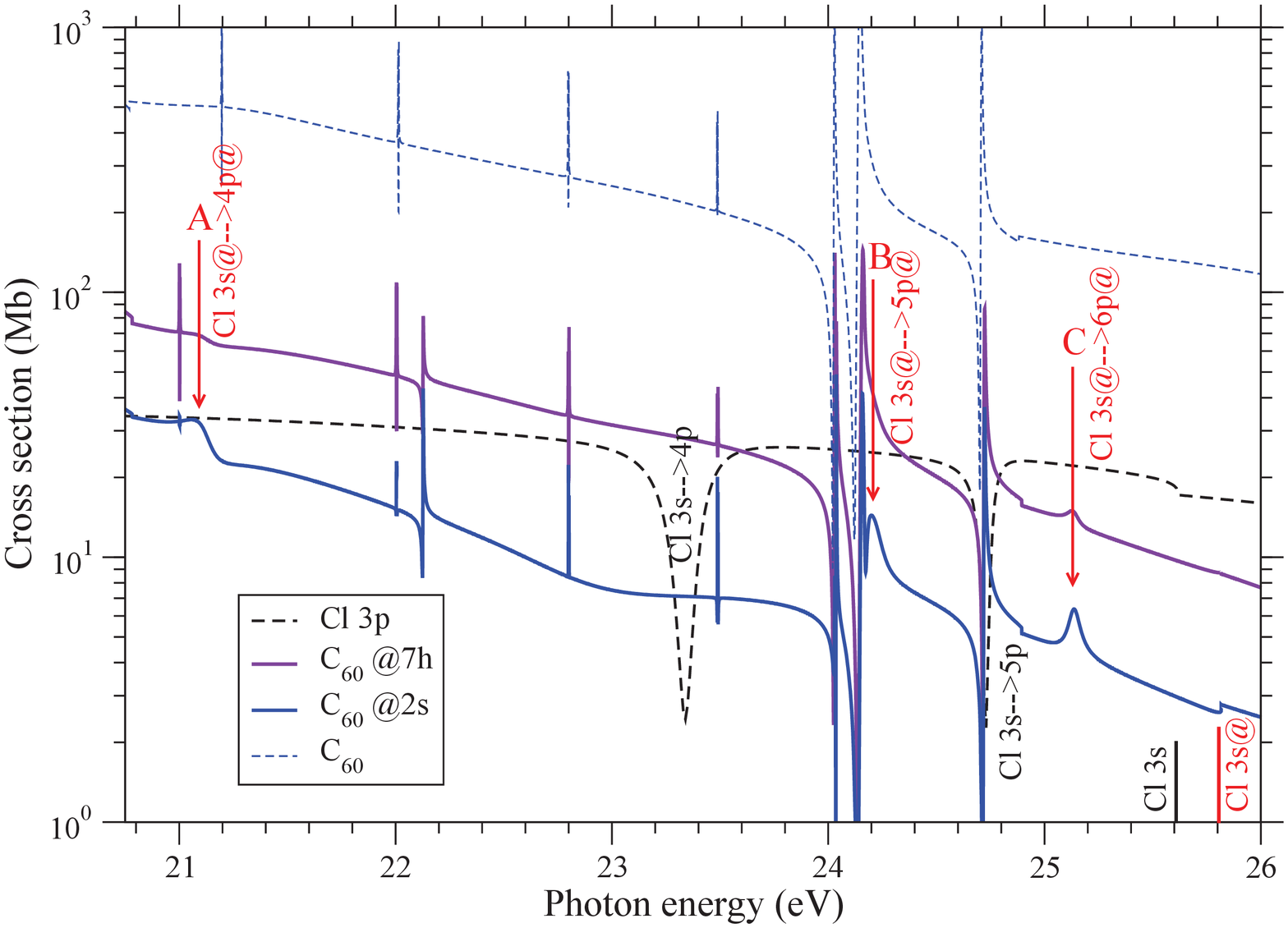,height=9.0cm,width=12.5cm,angle=0}}
	\caption{(Color online) Photoionization cross sections of free Cl $3p$ and empty $\ful$ compared with the results for $\ful$ $@7h$ and $@2s$ levels in Cl@$\ful$. Three Cl-to-$\ful$ ICD resonances (labeled as A,B,C) are identified in the $\ful$ @$7h$ and @$2s$ cross sections which can be compared with regular Auger resonances in free Cl $3p$.}
	\label{fig:figure1}
	\vspace{-0.35cm}
\end{figure}

Three such Cl-to-$\ful$ ICD resonances, corresponding to Cl $3s@ \rightarrow \eta p@$ with $\eta = 4,5,6$ (labeled as A, B and C), are seen in $\ful$ $@7h$ (HOMO) and $\ful$ $@2s$ cross sections in Fig.\,1. Note that these resonance features in $@2s$ are more prominent due to relatively smaller values of non-resonant background of the $@2s$ cross section. Also shown are the corresponding Auger resonances in free Cl $3p$ cross section from the decay of the first two $3s$ excitations which show clear Fano window-shape due to the higher background $3p@$ continuum transition strength. In comparison, the corresponding ICD resonances show dramatically different, small, peak-type shapes, indicating lower continuum transition strengths, besides the expected energy red-shifts, owing to the smaller binding energy of confined Cl $3s$@. These Cl-to-$\ful$ ICD resonances are qualitatively similar to those of Ar-to-$\ful$ found earlier~\cite{javani2014-rhaicd}, albeit with expected energy offsets. The remaining resonances in $@7h$ and $@2s$ cross sections in Fig.\,1 are from Auger decays of $\ful$ inner holes and are almost stable in their energies as can be seen by comparing with the empty $\ful$ total cross section (shown).

\subsection{$\ful$-to-Cl ICD resonances}

A coupled-channel representation of the matrix element like \eq{dc-mat-element1}, but to address the ICD resonances from the decay of $\ful$ inner excitations that appear in the Cl $3s@\rightarrow kp$ photoionization of Cl@$\ful$, can be written as,
\begin{eqnarray}\label{dc-mat-element2}
{M}^{\mbox{\scriptsize d-c}}_{3s@}(E) &=& \displaystyle\sum_{@n\ell} \sum_{\eta\lambda} \frac{\langle\psi_{@n\ell\rightarrow @\eta\lambda}|\frac{1}{|{\bf r}_1-{\bf r}_2|}
|\psi_{3s@\rightarrow kp}(E)\rangle}{E-E_{@n\ell\rightarrow\eta\lambda}} {\cal D}_{@n\ell\rightarrow @\eta\lambda}.
\end{eqnarray}
Since a number of inner $\ful$ vacancies can be produced that are degenerate with Cl $3s@$ ionization, two sums have been introduced. 
\begin{figure}
\vskip 0.7 cm
	\centerline{\psfig{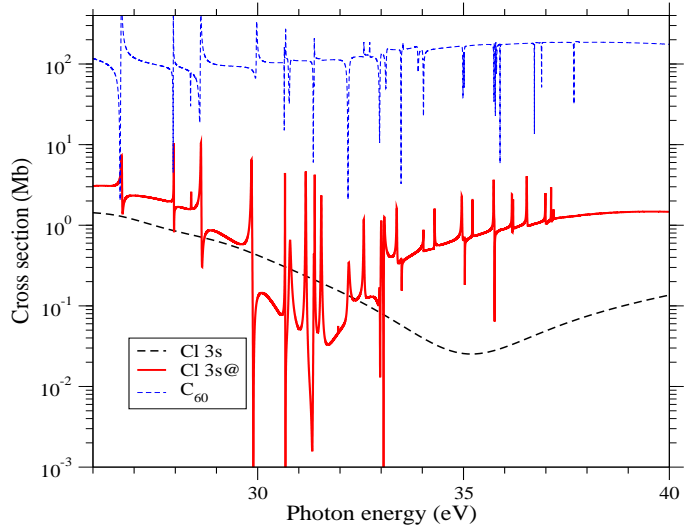}}
	\caption{(Color online) Cross sections of free Cl $3s$ subshell and $3s@$ of Cl@$\ful$ are compared. The total cross section of empty $\ful$ is also presented.}
	\label{fig:figure2}
	\vspace{-0.35cm}
\end{figure}

$\ful$-to-Cl ICD resonances are displayed in Fig.\,2 for $3s@$ photoionization of Cl. Note that the cross section is highly structured with the resonances when compared to the smooth $3s$ cross section of free Cl (shown). As seen, the free Cl result in the current energy range does not include any regular Auger decay of atomic innershell vacancies, indicating that the ICD process completely dominates the vacancy decay. The resonances are strong and of varied shapes. Their narrow width owes to the $\ful$ excitations. Indeed, $\ful$ wavefunctions, atypical of cluster properties, are delocalized, spreading over a large volume (see Fig.\,3). Since the autionization rate involves the matrix element of $1/r_{12}$ [\eq{dc-mat-element2}], spread-out wavefunctions translate to a decrease in the value of the matrix elements. 

Fig.\,2 also shows that for Cl $3s@$ the Cooper minimum, seen in the non-resonant background values of the curve, moves lower in energy to about 32 eV from its positions of 35 eV in free Cl. This shift is a consequence of the atom-$\ful$ dynamical coupling of Cl $ns@\rightarrow kp$ ionization channel with a host of $\ful$ continuum channels and was earlier noted for confined Ar and Kr as well~\cite{de2016}. This coupling is included in ${M}^{\mbox{\scriptsize c-c}}$ in \eq{gen-mat-element}. A comparison with the resonances (Auger) in the empty $\ful$ cross section (shown) indicates a general energy correspondence between Auger and ICD features, although there appear a rather dramatic shape alterations, in particular, at higher energies. The overall behavior of the ICD resonances is found very similar to the previous results for Ar $3s@$ and Kr $4s@$ caged in $\ful$.
\begin{figure}
\vskip 0.5 cm
	\centerline{\psfig{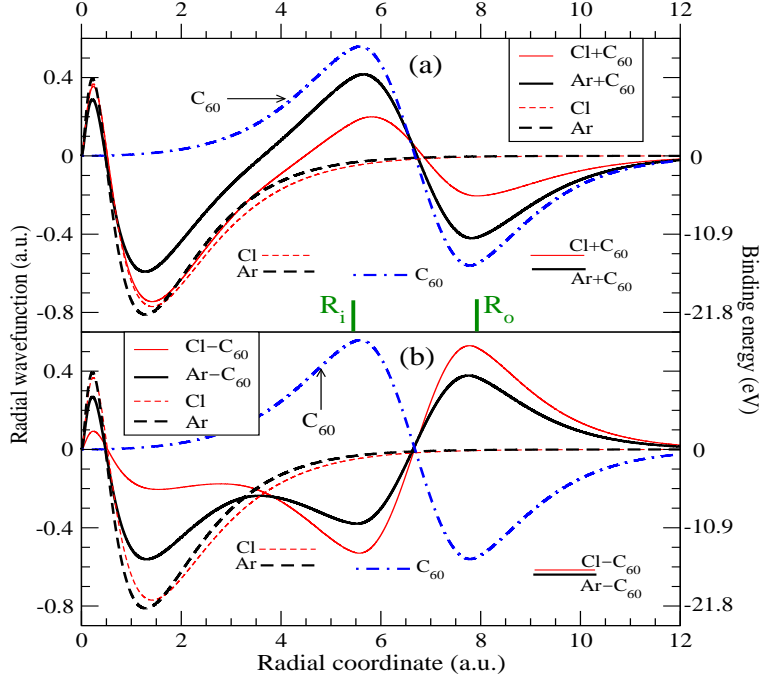}}
	\caption{(Color online) Radial symmetric (a) and antisymmetric (b) wavefunctions of Cl@$\ful$ \textit{versus} Ar@$\ful$. Wavefunctions of participant levels of free systems are displayed. Relevant binding energies in eV (scaled in opposite $y$-axis) are also graphed. The inner ($R_i$) and outer ($R_o$) radii of the $\ful$ shell are shown on panel (a).}
	\label{fig:figure3}
	\vspace{-0.1cm}
\end{figure}

\subsection{Hybrid Auger-ICD resonances}

About an equal share of mixing in ground state hybridization between valence $3p$ orbital for Ar with $\ful$ $3p$ was earlier found in Ar@$\ful$~\cite{javani2014-rhaicd}. In fact, the hybridization gap of 1.52 eV between (Ar+$\ful$) and (Ar$-\ful$) hybrid levels in that earlier calculation was in good agreement with the measured value of 1.6$\pm$0.2 eV \cite{morscher2010strong}. This hybridization in Cl@$\ful$ forming the symmetric and antisymmetric mixing similar to the bonding and antibonding states in molecules or dimers can be given as,
\begin{equation}\label{bound-hyb}
\mbox{Cl}3p\!\pm\!\ful 3p = |\phi_\pm\rangle = \sqrt{\alpha}|\phi_{3p \scriptsize{\mbox{Cl}}}\rangle \!\pm\! \sqrt{1-\alpha}|\phi_{3p \scriptsize{\mbox{C}_{60}}}\rangle.
\end{equation} 
A comparison in Fig.\,3 clearly shows somewhat weakened hybridization in Cl@$\ful$ versus Ar@$\ful$ which is primarily due to higher Cl $3p$ binding energy leading to a larger energy gap between this level with $\ful$ $3p$. Both free and hybrid radial wavefunctions and corresponding energies are shown in Fig.\,3 for a comparison between Cl@$\ful$ and Ar@$\ful$ ground state properties.  

Following \eq{bound-hyb}, the hybridization of the continuum channels assumes the form
\begin{equation}\label{channel-hyb}
|\psi_\pm\rangle = \sqrt{\alpha}|\psi_{3p@ \scriptsize{\mbox{Cl}}\rightarrow ks,d}\rangle \pm \sqrt{1-\alpha}|\psi_{@3p \scriptsize{\mbox{C}_{60}}\rightarrow ks,d}\rangle,
\end{equation}
where $\psi_{3p@ \scriptsize{\mbox{Cl}}\rightarrow ks,d}$ and $\psi_{@3p \scriptsize{\mbox{C}_{60}}\rightarrow ks,d}$ are the wavefunctions of the channels. Using \eq{channel-hyb} and recognizing that the overlap between a pure Cl and a pure $\ful$ bound state is negligible (see Fig.\,3), we may separate the atomic and fullerene regions of integration to write the matrix elements ${M}^{d-c}$ for emissions from hybrid levels as
\begin{eqnarray}\label{dc-mat-element3}
{M}^{\mbox{\scriptsize d-c}}_\pm (E) &=& \displaystyle\sum_{n\ell} \sum_{\eta\lambda}\left[\sqrt{\alpha}\frac{\langle\psi_{n\ell\rightarrow\eta\lambda}|\frac{1}{|{\bf r}_{\pm}-{\bf r}_{n\ell}|}
|\psi_{3p@ \scriptsize{\mbox{Cl}}\rightarrow ks,d}(E)\rangle}{E-E_{n\ell\rightarrow\eta\lambda}}\right. \nonumber \\
               \!\!&\!\!\pm\!&\!\left. \sqrt{1-\alpha} \frac{\langle\psi_{n\ell\rightarrow\eta\lambda}|\frac{1}{|{\bf r}_{\pm}-{\bf r}_{n\ell}|}
|\psi_{@3p \scriptsize{\mbox{C}_{60}}\rightarrow ks,d}(E)\rangle}{E-E_{n\ell\rightarrow\eta\lambda}} \right]\!\!{\cal D}_{n\ell\rightarrow\eta\lambda}. \nonumber \\
\end{eqnarray}

It is now straightforward to understand from \eq{dc-mat-element3} that if the inner vacancy corresponding to $n\ell\rightarrow\eta\lambda$ is located at Cl, then the first term in \eq{dc-mat-element3} denotes the decay through the Cl continuum, like in the Auger decay, and the second term will embody the decay through the $\ful$ continuum, like the ICD. This will result in resonant hybrid Auger-ICD (RHA-ICD) features from coherence in a outer hybrid level cross section driven by a Cl hole. These resonances for both hybrid levels of Cl@$\ful$ are presented in the top panel of of Fig.\,4 and labeled as A, B and C for three $3s@\rightarrow np@$ excitations. On the other hand, the original vacancy at $\ful$ in \eq{dc-mat-element3} will produce coherent RHA-ICD features initiated by a $\ful$ hole. All remaining resonances in Fig.\,4 (top) for hybrid states are from this latter category. We note in Fig.\,4 (top) that the resonances A-C in both Cl$\pm \ful$ $3p$ cross sections feature window-type shapes, like the free Cl Auger resonances (shown), generally suggesting strong continuum transitions. These RHA-ICD window resonances may likely be observed in the experiment due to their broad widths. However, we further note that these resonance shapes are significantly stronger for the symmetric Cl$+\ful$ level. The generic shapes of RHA-ICD structures from $\ful$ hole-decays, on the other hand, are seen to be non-window type, while they are substantially more prominent in the antisymmetric Cl$- \ful$ emission. The cause of these disparities is discussed below. 
\begin{figure}
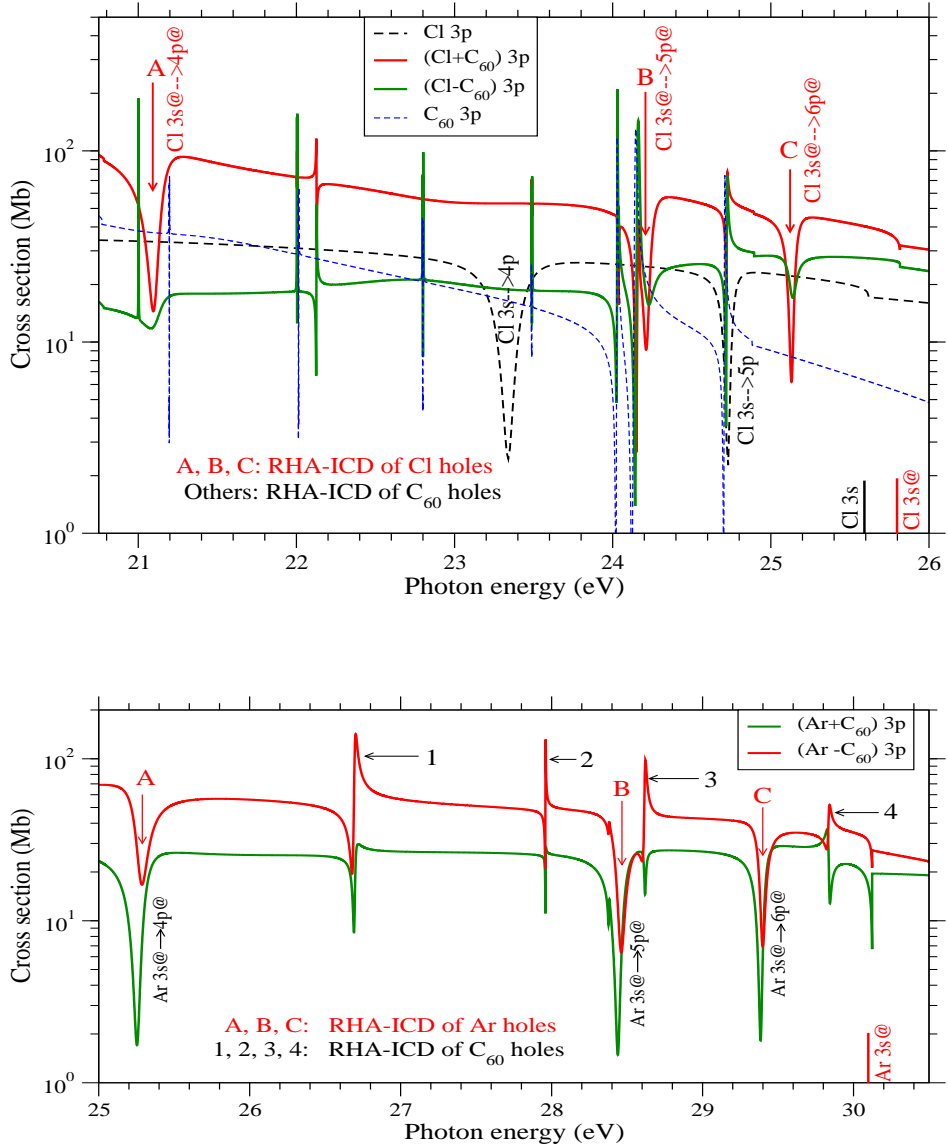

\vskip 1.0 cm
\centerline{\psfig{figure=fig4a.eps,height=8.0cm,width=12.5cm,angle=0}}
\vskip 1.2 cm
\centerline{\psfig{figure=fig4b.eps,height=6.0cm,width=12.5cm,angle=0}}

\caption{(Color online) (Top) Photoionization cross sections of free Cl $3p$ and $\ful$ $3p$ levels are compared with those of their hybrid pair. (Bottom) Published~\cite{javani2014-rhaicd} Ar@$\ful$ results for hybrid levels.}
\label{fig:figure4}
\end{figure}

The novel RHA-ICD features were originally predicted for Ar@$\ful$~\cite{javani2014-rhaicd} that are also presented in Fig.\,4 (bottom) to draw insights from comparisons. As seen for Ar@$\ful$, for each hybrid level cross section these resonances initiated by creation of Ar inner $3s$ holes (A-C) versus by creation of $\ful$ inner holes (1-4) are of approximately similar strengths. This is a direct consequence of a similar degree of mixing between Ar and $\ful$ characters in Ar$\pm\ful$ levels (see Fig.\,3) -- a fact that generates similar magnitudes of overlaps in the Coulomb coupling matrix elements in \eq{dc-mat-element3}. However, as noted before, the mixing reduces in Cl@$\ful$. As seen in Fig.\,3, while the Cl$+\ful$ level contains more Cl character, Cl$-\ful$ has more $\ful$ character. This fact, through the coupling matrix elements in \eq{dc-mat-element3}, translates into stronger RHA-ICD structures, A-C, for the decay of Cl holes in Cl$+\ful$ than in Cl$-\ful$. Likewise, this same fact is also responsible for the opposite behavior for $\ful$ hole decay, that is, the corresponding stronger structures in Cl$-\ful$ versus in Cl$+\ful$. 

\subsection{ETMD admixture}
In an endofullerene system the excited state for the precursor excitation resulting in an Auger decay or an ICD can itself be a state which is a hybrid of an atomic and a $\ful$  excited state of same angular momentum character. Here the excited states in question include $\eta p$@, @$\eta\lambda$ and $\eta\lambda$ in, respectively, \eq{dc-mat-element1}, \ref{dc-mat-element2} and \ref{dc-mat-element3}. This suggests that the part of the excited probability that will transfer to the other system can fall back in to the vacancy to initiate a resonant Auger decay or an ICD or a hybridized Auger-ICD. Furthermore, a very careful comparison between the ICD related results with corresponding empty $\ful$ results in Figs.\,(2) and (4) reveals a very few extra resonances in the ICD curves. These are present owing to the additional excited states in the spectrum of the whole compound, since it now also includes the empty states of the caged atom that can be transfer-excited by a $\ful$ electron. This clearly suggests that some of the resonances of both ICD and Auger nature presented here may incorporate coherent mixing with ETMD amplitudes. Of course, these contributions are hard to separate. A detailed discussion about such coherent combination of resonant ICD and ETMD mechanisms, but for a specific case of $\ful$-to-Ar ICD, was given elsewhere~\cite{de2016}. 

\section{Conclusion}

We present results of various kinds of single-electron ICD-type resonances in the photoemission of the Cl@$\ful$ molecule. This is the first ICD study of an endofullerene with a one-vacancy, open-shell atom inside. The calculation is carried out in a jellium-based linear-response time-dependent density functional framework that has previous success. The study includes resonant decays of Cl ($\ful$) innershell excitation vacancies degenerate with $\ful$ (Cl) outershell ionization vacancies. A uniquely different class of resonant features decaying into atom-fullerene hybrid final state vacancies has also been presented which arises from the interference of the Auger channel with an intrinsically connected ICD channel. These resonances are found to be remarkably strong, and the ones initiated by atomic excitations are quite broad. Hence they are likely experimentally measurable, allowing a powerful access to ICD dynamics. Furthermore, the hybridized character of some of the excited states of the compound points to a coherence of ICD with the ETMD process. Such ICD-ETMD coherences should be abundant -- all it would require is that both the fullerene and the trapped atom or molecule have dipole-allowed excited states of the same (angular momentum) symmetry so they can hybridize. Some of these resonances may also be amenable to being probed experimentally. Although the present calculation only includes {\em participant} RICDs, where the precursor hole is filled by the excited electron itself, it is of great interest to access the influence of {\em spectator} processes; these could significantly affect the situation and certainly need study. Based upon our explanation of the details of multicenter decay, the resonant ICD predicted here is expected to be a strong process in general for any atom or molecule encaged in any fullerene, in any position, central or not.

With the contemporary focus~\cite{dixit2013timedelay} on photoemission phase and time delay studies by interferometric metrology~\cite{kotur2016}, particularly at Fano window resonances for Ar~\cite{cirelle2018}, we hope that the current results will stimulate similar ultrafast spectroscopic studies of ICD resonances. And to that end, our future research outlook includes investigations of Wigner type intrinsic time delays of these various resonant ICD emissions.  
\\
\\
\\ 
{\large \bf Acknowledgment}\\
The work is supported by the National Science Foundation grant PHY-1806206 and the US Department of Energy, Office of Science, Basic Energy Sciences under Grant No. DE-FG02-03ER15428.

\end{document}